# MAXIMUM LIKELIHOOD ESTIMATES UNDER K-ALLELE MODELS WITH SELECTION CAN BE NUMERICALLY UNSTABLE

By Erkan Ozge Buzbas[1] and Paul Joyce[2]

*University of Idaho*

The stationary distribution of allele frequencies under a variety of Wright–Fisher $k$-allele models with selection and parent independent mutation is well studied. However, the statistical properties of maximum likelihood estimates of parameters under these models are not well understood. Under each of these models there is a point in data space which carries the strongest possible signal for selection, yet, at this point, the likelihood is unbounded. This result remains valid even if all of the mutation parameters are assumed to be known. Therefore, standard simulation approaches used to approximate the sampling distribution of the maximum likelihood estimate produce numerically unstable results in the presence of substantial selection. We describe the Bayesian alternative where the posterior distribution tends to produce more accurate and reliable interval estimates for the selection intensity at a locus.

**1. Introduction.** We begin by introducing a certain amount of terminology from population genetics. Within each living cell are a certain fixed number of *chromosomes*, threadlike objects that govern the inheritable characteristics of an organism. At certain positions, or *loci*, on the chromosomes, are *genes*, the fundamental units of heredity. At each locus there are several alternative types of genes or *alleles*. A *diploid* organism has chromosomes that occur in homologous pairs. The unordered pair of genes situated at the same locus of the homologous pair is called a *genotype*. Thus, if there are $k$ alleles, $A_1, A_2, \ldots, A_k$ at a given locus, then there are $k(k+1)/2$ possible genotypes, $(A_i A_j)$, $1 \leq i \leq j \leq k$. The *fitness* of a genotype $(A_i A_j)$ is determined by the reproductive success of individuals carrying that genotype.

Received July 2008; revised January 2009.

[1]Supported both by NSF Grant DEB-0515738 and NIH Grant P20 RR016454 from the INBRE Program of the National Center for Research Resources.

[2]Supported in part by NSF Grant DEB-0515738.

Understanding the evolutionary forces that shape the patterns of observed genetic diversity is central to population genetics. Over evolutionary time, genotypes with higher fitness tend to drive out those with lower fitness, thus reducing the genetic diversity of a population with respect to the particular locus under selection. However, there are selective mechanisms that actually promote genetic diversity. A simple such mechanism called heterozygote advantage assumes that carrying two variant copies of a gene at a locus (heterozygote) leads to higher fitness in comparison to carrying the same copy of the gene (homozygote). For example, individuals suffering from sickle cell anemia carry two copies of a disease gene. On the other hand, individuals that carry two copies of the healthy gene are susceptible to malaria. In regions with high incidence of malaria, individuals carrying one copy of the healthy gene and one copy of the diseased gene have higher fitness because they suffer only mild symptoms of anemia while also being resistant to malaria [Allison (1956), Cavalli–Sforza and Bodmer (1971), Harding and Griffiths (1997)]. As an infectious disease, malaria is a major health threat to human populations, affecting approximately 515 million people globally [Snow et al. (2005)].

For a population with allele frequencies $x_1, x_2, \ldots, x_k$, where $\sum_{i=1}^{k} x_i = 1$, we define the homozygosity to be $h = \sum_{i=1}^{k} x_i^2$, which is the probability that a sample of size two will produce two genes with the same allele. A population under the influence of heterozygote advantage would likely have low homozygosity. The lowest possible value occurs when all of the allele frequencies are equal ($x_i = 1/k$ for all $i$). Thus, low homozygosity might suggest that a high allelic diversity in a population is explained by heterozygote advantage. If all genotypes have equal fitness, we call the locus neutral. In this case, all of the genetic diversity is produced by mutation from one allele to another.

While the heterozygote advantage model assumes a diploid population, it is mathematically equivalent to a frequency dependent selection model which can apply to haploid organisms [Neuhauser (1999)]. One form of frequency dependence implies that high frequency alleles are at a selective disadvantage relative to alleles at low frequency. An example of this regime comes from allele frequencies from a bacteria that causes Lyme disease (*Borrelia burgdorferi*) [Donnelly, Nordborg and Joyce (2001)]. The data [previously published in Qiu et al. (1997)] consist of four alleles with frequencies $\mathbf{x}' = (0.103, 0.375, 0.270, 0.252)$. The observed homozygosity is $h = 0.288$, relatively close to the minimum 0.25 under $k = 4$. Under the neutral model assumption, mutations that occurred in the distant past would correspond to high frequency alleles and more recent mutations would give rise to low frequency alleles. So under neutrality one might expect a few alleles in high frequency and most alleles in low frequency, corresponding to relatively high homozygosity. At first glance, the homozygosity in the above data appears to be too low to be explained by neutrality. Watterson derived the distribution of the homozygosity statistic under the assumption of neutrality which lead to one of the most common tests to distinguish departures from neutrality [Watterson (1977)]. However, with modern computational methods, we are now in a position to go beyond simply determining whether neutrality is feasible. We can now develop likelihood based inference methods for precise alternative models that incorporate selection and estimate the strength of selection. The alternative models presented here are called the Wright–Fisher $k$-allele models with selection [Wright (1949)]. These classical models have a rich mathematical theory and long history in population genetics [see Ewens (2004) for background], mainly because they provided theoretical insight into the dynamics of genes evolving under selection. In the data rich world that population genetics finds itself today, there is a renewed interest in these models as useful tools to draw inferences on selection at genetic loci of interest.

A brief description of the Wright–Fisher process is as follows. Consider tracking a population of $2N$ genes over many generations. A gene can be one of $k$ possible alleles. Generations are non overlapping and the probability of sampling genotype $(A_i A_j)$ is proportional to its fitness, $w_{ij} = 1 - s_{ij}$. To obtain the next generation, $2N$ pairs of genes are sampled with replacement. A randomly chosen allele within each sampled pair mutates to $A_i$ with probability $u_i$, independent of the parent's type. A standard diffusion argument [Ewens (2004)] based on the Markov chain of allele frequencies generates the stationary distribution

$$(1) \qquad f_{\text{Sel}}(\mathbf{x} | \boldsymbol{\theta}, \boldsymbol{\Sigma}) = \frac{e^{-\mathbf{x}' \boldsymbol{\Sigma} \mathbf{x}}}{E_{\text{Neut}}(e^{-\mathbf{X}' \boldsymbol{\Sigma} \mathbf{X}})} f_{\text{Neut}}(\mathbf{x} | \boldsymbol{\theta}),$$

where $\mathbf{x}$ is a $(k \times 1)$ column vector of allele frequencies in the population, subject to the condition $\sum_{i=1}^{k} x_i = 1$, whose transpose is the row vector $\mathbf{x}' = (x_1, \ldots, x_k)$. We have $\boldsymbol{\theta}' = (\theta_1, \ldots, \theta_k)$, where $\theta_i = 4 N u_i$, is the scaled mutation parameter for type $i$ and $\boldsymbol{\Sigma} = (\sigma_{ij})$ with $\sigma_{ij} = 2 N s_{ij}$, the $(k \times k)$ symmetric matrix of scaled selection parameters. The probability density function for allele frequencies under neutrality, $f_{\text{Neut}}(\mathbf{x} | \boldsymbol{\theta})$, is given by the familiar Dirichlet distribution

$$(2) \qquad f_{\text{Neut}}(\mathbf{x} | \boldsymbol{\theta}) = \frac{\Gamma(\theta_1 + \theta_2 + \cdots + \theta_k)}{\Gamma(\theta_1) \Gamma(\theta_2) \cdots \Gamma(\theta_k)} x_1^{\theta_1 - 1} x_2^{\theta_2 - 1} \cdots x_k^{\theta_k - 1}.$$

We will use the notation $E_{\text{Neut}}(\cdot)$ to represent expectation with respect to the neutral density given by equation (2), and $E_{\text{Sel}}(\cdot | \boldsymbol{\Sigma})$ to denote expectation with respect to the density that incorporates selection given by equation (1).

The symmetric selective overdominance model is a special case of equation (1), obtained by setting $w_{ij} = 1$, that is $(s_{ij} = 0)$ for heterozygotes $(i \neq j)$ and $w_{ii} = 1 - s$, $(s_{ii} = s > 0)$ for homozygotes regardless of the specific type and assuming symmetric mutation $\theta_i = \theta/k$, for all $i$. Under these assumptions, the matrix of selection parameters reduces to a diagonal matrix with equal elements, $\mathbf{\Sigma} = \sigma \mathbf{I}_k$, where $\mathbf{I}_k$ is the $k$ dimensional identity matrix, $\sigma = 2Ns$, and the mutation parameter is $\theta = 4Nu$. Therefore, $\mathbf{X}'\mathbf{\Sigma}\mathbf{X} = \sigma \sum_{i=1}^{k} X_i^2$. Substituting appropriately into equation (1) gives

$$(3) \qquad f_{\text{Sel}}(\mathbf{x}|\theta, \sigma) = \frac{e^{-\sigma \sum_{i=1}^{k} x_i^2}}{E_{\text{Neut}}(e^{-\sigma \sum_{i=1}^{k} X_i^2})} \frac{\Gamma(\theta)}{(\Gamma(\theta/k))^k} (x_1 x_2 \cdots x_k)^{\theta/k - 1}.$$

Despite the wide ranging applications of $k$-allele models, the statistical properties of estimators are not well understood. This article aims to clarify inference problems associated with estimators of the selection intensity and presents correct frequentist and Bayesian methods for inference under $k$-allele models. Theoretical results describing the large variability in the sampling distribution of maximum likelihood estimates (MLEs), arising in the analysis of $k$-allele models with selection in general and of symmetric overdominance in particular, are given. The likelihood of allele frequencies under the stationary distribution of the diffusion limit of a Wright–Fisher population is examined and the existence of a numerical instability associated with MLEs for certain population compositions is established. Numerical instability of MLEs occurs under what would normally be considered ideal conditions. These conditions include the assumption that all the mutation parameters are either known or can be estimated without error. Also, the allele frequencies of the entire population are assumed to be observed, rather than the usual assumption that the data are viewed as a random sample from the population. Even under these idealized conditions, it is shown that when the data carry a strong signal for selection, parametric bootstrap is inaccurate and unreliable for assessing the strength of selection.

In Section 2 the theoretical basis for the instability behavior is formalized by a theorem: under $k$-allele models with selection, there is a singularity point on the allele frequency space where the likelihood is unbounded. A corollary associated with the theorem is also presented: under the symmetric overdominance model, the above mentioned singularity arises when all the alleles have equal frequencies. This is identified as a perfectly heterozygous population. This result is surprising since it suggests that appreciable information in the data about selection yield poor estimates and MLEs with a parametric bootstrap approach cannot be used effectively to estimate the strength of selection. Therefore, our findings highlight the limitations of using the sampling distribution of the MLEs for inference under $k$-allele models with selection.

Section 3 exploits a monotonicity argument to show a correct frequentist approach to the problem as well as a Bayesian method as an alternative to parametric bootstrap. Data from a Killer-cell immunoglobulin-like receptor

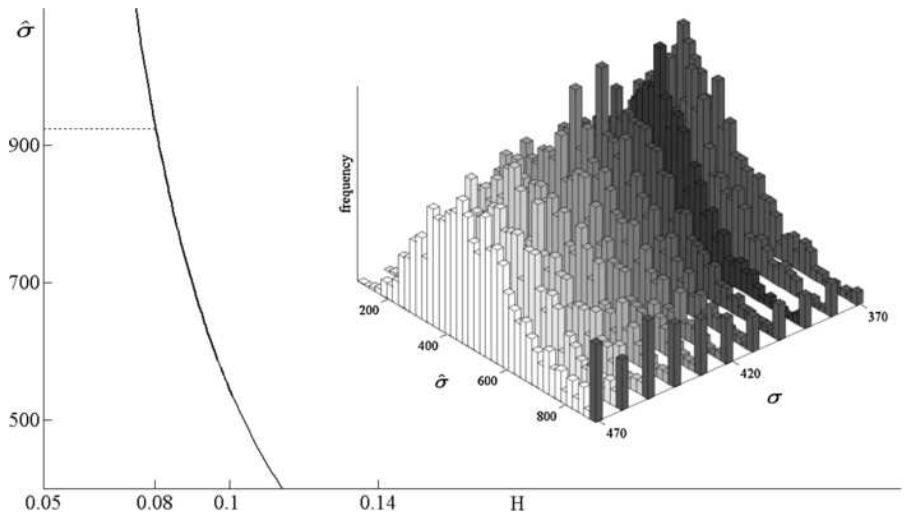

Fig. 1. *Left: The distribution of the maximum likelihood estimates as a function of population homozygosity based on equation (3) considering data space that can arise under $k = 20$ on a grid. $\hat{\sigma}$ asymptotes as h approaches to $1/k = 0.05$. Right: Heavy tailed sampling distributions of $\hat{\sigma}$, each based on 1000 simulated data sets with $\theta = 5$.*

(KIR) locus and the Lyme disease bacteria locus discussed above are used to demonstrate the methods in Section 4. A comparison of the heterozygote advantage with another selective mechanism of importance, the homozygote advantage, is given in Section 5, to emphasize that the effect of instability in MLEs may be different under different selective scenarios.

**2. Instability of the MLE.** A useful summary of the population composition under the overdominance model is the homozygosity statistic $H = \sum_i X_i^2$, which is sufficient for $\sigma$ given $\theta$ and the MLE $\hat{\sigma}(h)$, is a decreasing function of $h = \sum_i x_i^2$.

The signal for selection is strong when the population homozygosity is small (Figure 1, left). However, the high rate of divergence and unboundedness of $\hat{\sigma}(h)$ as $h$ approaches to its minimum value $1/k$ is unexpected. Small perturbations of allele frequencies have a drastic effect on point estimates, particularly for small $h$. For instance, for a highly polymorphic locus with $k = 20$ possible alleles, where $1/k = 0.05$, a 38% decrease in homozygosity ($h = 0.13$ to $h = 0.08$) corresponds to an approximate 300% increase in $\hat{\sigma}(h)$; $[\hat{\sigma}(0.13) \approx 350$, whereas $\hat{\sigma}(0.08) > 900]$.

We use the numerical methods of [Genz and Joyce (2003), Joyce, Genz and Buzbas (2008)] to generate data from the density (3) and to evaluate the likelihood based on it. These simulated data are then used to estimate the error of $\hat{\sigma}$. Assume we condition on $\theta$, so that the sole focus of inference is on $\sigma$. When a simulated sample with $h \approx (1/k)$ is drawn, the resulting $\hat{\sigma}$ is large. Thus, such samples contribute to a heavy right tail for the distribution of $\hat{\sigma}$. Sampling distributions generated under strong selection clearly reveal the substantial effect of the heavy tail on the estimation of $\sigma$ (Figure 1, right). Estimates from data sets generated under a variety of $(\sigma, \theta, k)$ values show that the heavy tail in the distribution of $\hat{\sigma}$ is a persistent feature. In fact, *all* $k$-allele models have a singularity in data space. Theorem 1 below gives a precise statement of this general phenomenon.

THEOREM 1. *Consider the probability density function $f_{\text{Sel}}(\mathbf{x}|\boldsymbol{\theta}, \boldsymbol{\Sigma})$ defined by equation (1) that describes the distribution of allele frequencies at stationarity under the Wright–Fisher model with selection and parent independent mutation. There exists a vector of allele frequencies $\mathbf{x}^* = (x_1^*, \ldots, x_k^*)'$, where $f_{\text{Sel}}(\mathbf{x}^*|\boldsymbol{\theta}, \boldsymbol{\Sigma})$ is unbounded as a function of $\boldsymbol{\Sigma}$ regardless of $\theta$.*

The proof of Theorem 1 is given in Appendix A, where we show that the point $\mathbf{x}^*$ where $f_{\text{Sel}}(\mathbf{x}^*|\boldsymbol{\theta}, \boldsymbol{\Sigma})$ is unbounded as a function of $\boldsymbol{\Sigma}$ is found by minimizing the quantity $\sum_{i,j} \sigma_{ij} x_i x_j$ subject to the constraint $\sum_i x_i = 1$. If the mean fitness of the population is defined by $\bar{w} = \sum_{i,j} w_{ij} x_i x_j$, we have $\sum_{i,j} \sigma_{ij} x_i x_j = 2N(1 - \bar{w})$. Therefore, $\mathbf{x}^*$ is a point in the data space where $\bar{w}$ is optimal.

The symmetric overdominance model, which has a considerably simpler matrix of selection parameters, allows for further results. In this case, both the limiting value for $\hat{\sigma}$ and the point $\mathbf{x}^*$ at which the likelihood is unbounded can be established.

COROLLARY 1. *Consider the probability density function $f_{\text{Sel}}(\mathbf{x}|\theta, \sigma)$, defined by equation (3) that describes the distribution of allele frequencies at stationarity for the Wright–Fisher symmetric selective overdominance model with parent independent mutation. Let $\mathbf{x}^* = (1/k, \ldots, 1/k)'$.*

a. *If $\theta$ is assumed to be known, then, for all allele frequencies $\mathbf{x} \neq \mathbf{x}^*$, the maximum likelihood estimate for $\sigma$ is finite. Denote the MLE as a function of the homozygosity $h = \sum_{i=1}^k x_i^2$ by $\hat{\sigma}(h)$. Then,*

$$\text{(4)} \qquad\qquad \lim_{h \to (1/k)^+} \hat{\sigma}(h) = \infty.$$

b. *For all $\theta > 0$, if $(X_1, X_2, \ldots, X_k)$ has joint probability density given by equation (3), then $(X_1, X_2, \ldots, X_k)$ converges in probability to $(1/k, 1/k, \ldots, 1/k)$ as $\sigma \to \infty$.*

*(See Appendix B for proof).*

Part (b) of Corollary 1 provides some context for the ill behavior of the MLE for $\sigma$. If $\sigma$ is large, it is highly likely that the population frequencies are nearly equal. The asymptotic behavior of populations in the limit as $\sigma$ goes to infinity has been studied in other contexts by both Gillespie (1999) and Joyce, Krone and Kurtz (2003).

## 3. Methods to assess the error in estimates.

3.1. *Estimates using the monotonicity of homozygosity.* By exploiting the monotonicity of the distribution of the homozygosity statistic, $H$, we develop a frequentist approach to obtain interval estimates of selection intensity under the selective overdominance model [equation (3)] that does not rely on approximating the sampling distribution for the MLE $\hat{\sigma}$. Throughout we assume that $\theta$ is known and we drop the parameter $\theta$ for notational convenience when denoting the cumulative distribution function (cdf) for the homozygosity $H = \sum_{i=1}^{k} X_i^2$ such that we have

$$F_{\mathrm{H}}(h|\sigma) = P_{\mathrm{Sel}}(H \leq h|\theta, \sigma).$$

While the cdf is always a monotone function of $h$ for fixed $\sigma$, this particular cdf is also monotone with respect to the parameter $\sigma$ for fixed $h$. More precisely, we can state that as $\sigma$ increases, the probability of $H$ being smaller than a particular value $h$ increases. An exact confidence interval for $\sigma$ is produced using this monotonicity property in $\sigma$.

For a given confidence level $(1 - \alpha)$ and an observed homozygosity $H = h$, we choose $\hat{\sigma}_L$ and $\hat{\sigma}_U$ so that

(5) $$F_{\mathrm{H}}(h|\hat{\sigma}_L) = \alpha_1, \qquad F_{\mathrm{H}}(h|\hat{\sigma}_U) = 1 - \alpha_2,$$

where $\alpha = \alpha_1 + \alpha_2$. We interpret $\hat{\sigma}_L$ and $\hat{\sigma}_U$ as the smallest and largest values of $\sigma$ that supports the data. Since $F_{\mathrm{H}}(h|\sigma)$ is a monotone increasing function of $\sigma$, then $\hat{\sigma}_L \leq \sigma \leq \hat{\sigma}_U$ if and only if $F_H(h|\hat{\sigma}_L) \leq F_H(h|\sigma) \leq F_H(h|\hat{\sigma}_U)$, which implies $\alpha_1 \leq F_H(h|\sigma) \leq 1 - \alpha_2$. Therefore,

$$P_{\mathrm{Sel}}(\hat{\sigma}_L \leq \sigma \leq \hat{\sigma}_U) = P_{\mathrm{Sel}}(\alpha_1 \leq F_H(H|\sigma) \leq 1 - \alpha_2).$$

The standard theory shows that $F_{\mathrm{H}}(H|\sigma)$ is a uniformly distributed random variable on the interval $[0, 1]$ to give the result

$$P_{\mathrm{Sel}}(\hat{\sigma}_L \leq \sigma \leq \hat{\sigma}_U) = 1 - \alpha_1 - \alpha_2 = 1 - \alpha.$$

Therefore, $[\hat{\sigma}_L, \hat{\sigma}_U]$ is an *exact* $(1 - \alpha)$ level confidence interval.

However, when $\theta$ is unknown, the monotonicity of $H$ in $\sigma$ holds no more and only *confidence regions* are possible to obtain. Therefore, in applications where the variability in $\hat{\theta}$ is expected to be considerable and the joint estimation of $(\theta, \sigma)$ is required, the method is not very useful. Next, we turn to the Bayesian approach as an alternative that allows for marginalization in $\sigma$.

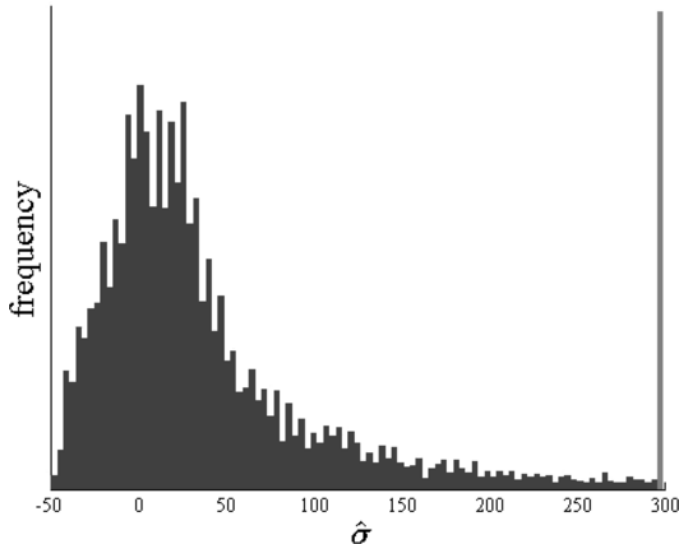

FIG. 2. *A parametric bootstrap sample of size* 10000 *for $\hat{\sigma}$ using the Lyme disease data. All values of $\hat{\sigma} > 300$ are plotted at* 300.

3.2. *Estimates based on the Bayesian methods.* Assuming independent uniform priors on $(\theta, \sigma)$, the joint posterior distribution of $(\theta, \sigma)$ is proportional to the likelihood,

$$(6) \qquad P_{\text{Sel}}(\theta, \sigma | \mathbf{x}) \propto \frac{e^{-\sigma \sum_{i=1}^{k} x_i^2}}{E_{\text{Neut}}(e^{-\sigma \sum_{i=1}^{k} X_i^2})}(x_1 x_2 \cdots x_k)^{\theta/k-1},$$

which can be sampled using a standard Markov Chain Monte Carlo approach. We sampled the joint posterior distribution of the parameters via an independent Metropolis–Hastings update [Metropolis et al. (1953), Hastings (1970)]. The posterior mode is found by numerically maximizing the joint distribution and 95% credible intervals are used as a measure of variability for $\sigma$.

Since both the posterior and the bootstrap are based on the likelihood, intuition suggests a similar problem of instability might arise in the Bayesian analysis. Examining the posterior sample of $\sigma$ (see examples in Section 4), we find that the Bayesian approach does not have the instability observed in the bootstrap. The reason is that, in contrast to the parametric bootstrap which generates data, posterior analysis works on *fixed* data. While each simulated data set in the bootstrap has a certain probability of falling into the instability region, this problem is avoided in posterior simulation, by sampling the parameter space rather than the data space.

**4. Examples.** In this section we present two data applications to compare the three methods discussed above for making inference on selection intensity: MLE-bootstrap, monotonicity and the Bayesian approach.

4.1. *Lyme disease data.* We revisit (see Section 1) the Lyme disease bacteria data. Recall the data consist of four alleles with frequencies $\mathbf{x}' = (0.103, 0.375, 0.270, 0.252)$. The observed homozygosity is $h = 0.288$ relatively close to the minimum 0.25 under $k = 4$. The MLEs for the mutation and selection parameters are $(\hat{\theta} = 4.8, \hat{\sigma} = 35.1)$ respectively. A parametric bootstrap with $(\hat{\theta}, \hat{\sigma}, k) = (4.8, 35.1, 4)$ admits poor precision for the estimated selection intensity as discussed in Section 2. Based on the simulated sampling distribution for $\hat{\sigma}$ (Figure 2), we get an estimated standard error of 176.4. The 2.5th percentile of the simulated sampling distribution of $\hat{\sigma}$ corresponds to 17.2 and the 97.5th percentile is 681.3. Therefore, an approximate 95% interval estimate based on the parametric bootstrap associated with $\hat{\sigma}$ is $(17.2, 681.3)$. Recall that the observed homozygosity is 0.228, but $P(H \geq 0.288 | \sigma = 681.2) < 0.001$, suggesting that the upper bound produced by the parametric bootstrap is far too conservative. Conversely, $P(H \leq 0.288 | \sigma = 17.25) = 0.354$, suggesting that the lower bound produced by the parametric bootstrap is too large to be reliable. Thus, the parametric bootstrap is both unreliable and inaccurate. Using the monotonicity of homozygosity with $\alpha_1 = \alpha_2 = 0.025$ produces an exact 95% confidence interval of $(-8, 105)$ for the Lyme disease data. The upper bound from this method performs much better than the bootstrap, as expected. The length of the exact confidence interval is over six times smaller than the length of the interval produced by the parametric bootstrap.

A 95% credible interval from the posterior simulation gives $(10.8, 124.9)$. The length of the interval produced by the parametric bootstrap is over six times larger than the interval estimate produced by the Bayesian method.

4.2. *KIR data.* Our second example is from a data set published in Norman (2004) on KIR genes. The data are from the United Kingdom population [see locus DL1/S1 from Table 2 in Norman (2004)]. The KIR are highly polymorphic genes coding for proteins on natural killer cells and they detect a specific Major histocompatibility complex Class I protein found on diseased natural killer cells. Observed levels of variability at these loci suggest that the heterozygote advantage mechanism is a good candidate to explain the variation in KIR genes. The population frequencies are given by $\mathbf{x}' = (0.22, 0.21, 0.17, 0.16, 0.15, 0.04, 0.03, 0.02)$. The homozygosity statistic is $h = 0.172$, again close to the minimum $h_{\min} = 0.125$ for $k = 8$. An approximate 95% bootstrap interval estimate for this data set is $(21.1, 396.4)$. Conditioning on $\hat{\theta}$, the interval estimate using the monotonicity argument with $\alpha_1 = \alpha_2 = 0.025$ is given as $(-10, 159)$ (Figure 3), which is less than half the length of the bootstrap interval.

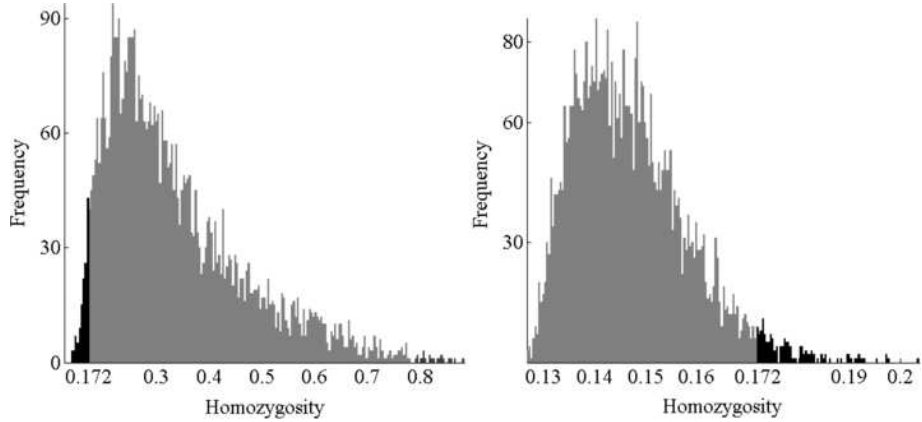

FIG. 3. *The empirical distributions of homozygosity under $\hat{\sigma}_L = -10$ (left) and $\hat{\sigma}_U = 159$ (right) for KIR data (h = 0.172). The shaded areas correspond to 2.5th and 97.5th percentiles.*

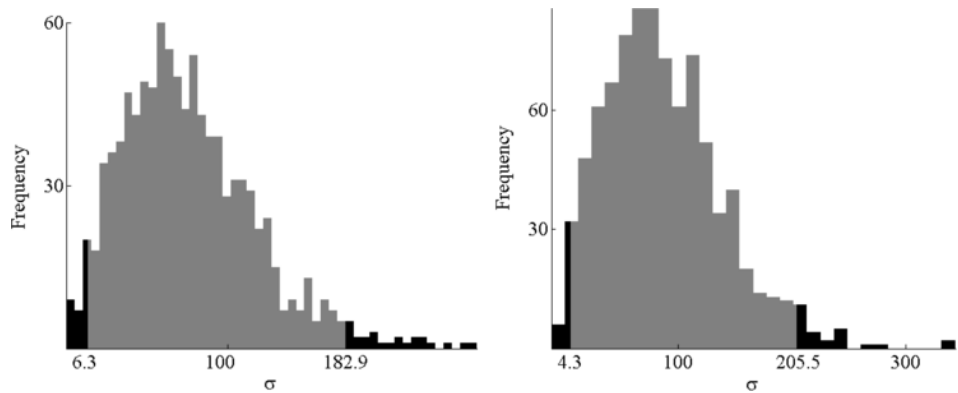

FIG. 4. *A sample from posterior distribution of $\sigma$ for the KIR data set using fixed $\theta$ (left) and the joint estimation (right). 95% credible interval limits are (6.3, 182.9) and (4.3, 205.5) (shades) respectively.*

Fixing $\theta$ at the posterior mode, the KIR data giving a 95% credibility interval for $\sigma$ is $(6.3, 182.9)$ (Figure 4). For comparison purposes a 95% credibility interval for $\sigma$, $(4.3, 205.5)$, is obtained by joint estimation of $\theta$ and $\sigma$ is also included in Figure 4. Not surprisingly, the variability in $\sigma$ increases when the uncertainty in $\theta$ is taken into account.

**5. Discussion.** Wright–Fisher $k$-allele models with selection provide a flexible framework for considering a wide array of biologically meaningful selective schemes. The impact of the instability on the MLEs can vary substantially depending on the particular selective scheme. In this section we describe an example using the symmetric homozygote advantage model. Our goal is to compare the homozygote and heterozygote advantage models in terms of the instability explained. The comparison is particularly insightful, as the two schemes have very different biological implications and yet there is a close connection between their parameterization.

In contrast to heterozygote advantage, homozygote advantage selects for genotypes with the same allelic types. The symmetric version can be obtained by letting the selection matrix $\mathbf{\Sigma} = -\sigma \mathbf{I}_k$, $\sigma > 0$, now denoting the relative selective advantage of homozygotes to heterozygotes. Note that the difference between this model and the heterozygote advantage consists only of switching the sign of the selection parameter. Therefore, both heterozygote and homozygote advantages can be accommodated in the same model by allowing $\sigma \in \mathbb{R}$, a useful property to compare the instability regions arising under two regimes. Simulated sampling distributions obtained under the homozygote advantage display a heavy left tail, reflecting the sign change in $\sigma$. In light of Theorem 1, which holds for all selective schemes, this result is not surprising. The existence of a singularity point for the homozygote advantage model is guaranteed. The strongest signal for selection under this model is at $H = 1$ (i.e., at maximum homozygosity) and, in fact, the same point turns out to be where $\hat{\sigma}$ is infinite since this point maximizes the mean fitness.

Interestingly, the effect of the instability on the variability of MLEs under the homozygote advantage is milder than the heterozygote advantage. The difference is explained by examining the populations contributing to the tail of interest in the corresponding bootstrap distributions (i.e., left and right tails for homozygote and heterozygote advantage models respectively). They have different probabilities of arising under the two cases. Let $0 < \varepsilon < 1 - 1/k$ and consider populations that are close to perfect homozygosity $(1 - \varepsilon < H < 1)$, and those that are the same distance from the perfect heterozygosity $(1/k < H < 1/k + \varepsilon)$. In the bootstrap procedure, whenever a generated data set falls in these regions, a large MLE for the selection intensity is obtained. However, the probability of drawing a population in the first set, $P_{\mathrm{Sel}}(1 - \varepsilon < H < 1 | \theta, -\sigma)$, is lower than the probability of drawing a population in the second set, $P_{\mathrm{Sel}}(1/k < H < 1/k + \varepsilon | \theta, \sigma)$, for a given (absolute) selection intensity (Figure 5). In other words, for a given number of simulated samples, the expected number of samples to hit the singularity region is larger under symmetric heterozygote advantage than under the homozygote advantage. Hence, the type of selective regime is an important factor when evaluating the effect of the instability on the confidence intervals. Unfortunately, in the important case of the heterozygote advantage model, the effect is pronounced.

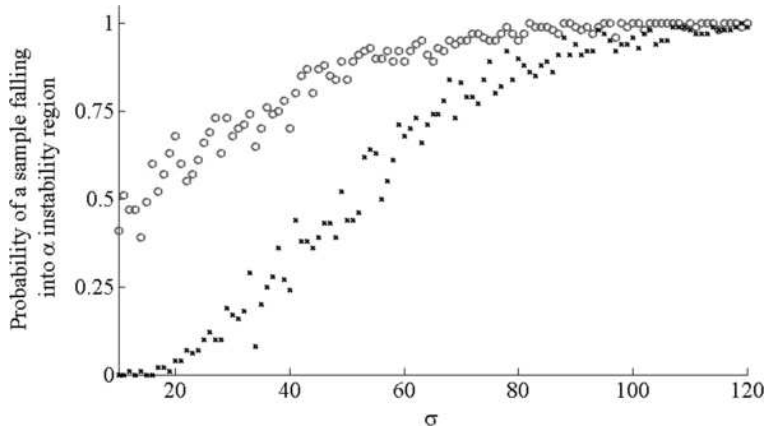

Fig. 5.  *The probability that a sample falls into the instability region for different selection intensities (as percent of generated samples with 1000 samples at each $\sigma$) under homozygote advantage (dots) and heterozygote advantage (open circles). The samples are generated under $k = 10$ and the instability region is chosen using $\varepsilon = 0.09$ units from perfect homozygosity (1, 0.91) and heterozygosity (0.1, 0.19) (see text).*

The ultimate goal for the use of methods discussed in this paper is to develop statistical methods that can be used to detect selection at multiple loci simultaneously under the $k$ allele models. Multiple loci data provide more information than single locus data, therefore, inference is expected to be more precise. The Bayesian methods gain a definite advantage of flexibility as the number of genetic loci, and thus the dimensionality of the problem, increases.

Modern population genetics using coalescent based methods to explain polymorphism data have been effectively used to understand genealogy and recombination [Fearnhead (2001), Nordborg (2000), Griffiths and Marjoram (1997), Padhukasahasram et al. (2008)] but have been less successful with selection. The computational burden for simulating and analyzing data under coalescent based methods with selection remains heavy. There is a renewed interest in diffusion approaches which provide an alternative framework to handle models with selection [Wakeley (2005)]. Furthermore, diffusion models are cornerstones of population genetics theory. Their relatively long history resulted in a variety of useful models to investigate selection other than the $k$-allele setup. An important task is to establish statistical properties of estimators and investigate the usefulness of different statistical paradigms under these models.

Finally, counterintuitive results presented in this paper point out that care should be exercised in method choice when making inference on selection under the class of models we presented. As emphasized above, realistic applications of the methods involve inference from multiple loci possibly with complex selective schemes. However, such setups are not ideal to investigate the statistical properties of the estimators. Because computationally intensive methods employed in analyzing them become less tractable, complex models tend to hide problems of the type discussed in this paper. Therefore, rigorous tests of the methods under the single locus case are essential to guarantee the legitimacy of inference on selection made by employing these methods.

## APPENDIX A: PROOF OF THEOREM 1

We begin by fixing $\boldsymbol{\theta}$ and $\boldsymbol{\Sigma}$, then finding a point in data space that produces the largest possible signal for selection. Data space is represented by the $k$ dimensional simplex defined by $\Delta_k = \{(x_1, x_2, \ldots, x_k) \mid \sum_{j=1}^k x_j = 1\}$. Since $\Delta_k$ is a compact set, there exists at least one point $\mathbf{x}^* \equiv \mathbf{x}^*(\boldsymbol{\Sigma}) \in \Delta_k$ where $e^{-\mathbf{x}\boldsymbol{\Sigma}\mathbf{x}'}$ is optimized. It follows from equation (1) that $\mathbf{x}^*$ is the point in data space that produces the strongest possible signal for selection. Note that $\mathbf{x}^*$ is obtained by minimizing the quadratic function $\mathbf{x}\boldsymbol{\Sigma}\mathbf{x}' = \sum_{ij} \sigma_{ij} x_i x_j$ subject to the constraint $\sum_i x_i = 1$. This implies

$$(7) \qquad \sum_{i,j} \sigma_{ij} x_i^* x_j^* \leq \sum_{i,j} \sigma_{ij} x_i x_j \qquad \forall \mathbf{x} \in \Delta_k.$$

We now turn to the alternative problem where we fix the data $\mathbf{x}$ and calculate the maximum likelihood estimate for $\boldsymbol{\Sigma}$ denoted by $\hat{\boldsymbol{\Sigma}}(\mathbf{x})$. Throughout we will assume that $\boldsymbol{\theta}$ is known. Since the likelihood [equation (1)] is a smooth function of the parameters, the standard calculus approach to optimization based on the derivative of the log likelihood is a valid method for finding the MLE. It follows from equation (1) that

$$
\begin{aligned}
(8) \qquad \frac{\partial}{\partial \sigma_{ij}} \ln f_{\mathrm{Sel}}(\mathbf{x}|\boldsymbol{\theta},\ \boldsymbol{\Sigma}) &= \frac{E_{\mathrm{Neut}}(X_i X_j e^{-\mathbf{X}\boldsymbol{\Sigma}\mathbf{X}'})}{E_{\mathrm{Neut}}(e^{-\mathbf{X}\boldsymbol{\Sigma}\mathbf{X}'})} - x_i x_j \\
&= E_{\mathrm{Sel}}(X_i X_j | \boldsymbol{\Sigma}) - x_i x_j.
\end{aligned}
$$

Therefore, to obtain the MLE for $\boldsymbol{\Sigma}$, denoted by $\hat{\boldsymbol{\Sigma}}(\mathbf{x})$, we set equation (8) equal to zero for each pair of indices $i, j$ and solve for $\hat{\boldsymbol{\Sigma}}(\mathbf{x})$. Thus, for a given data set $\mathbf{x}$, we have

$$(9) \qquad E_{\mathrm{Sel}}(X_i X_j | \hat{\boldsymbol{\Sigma}}(\mathbf{x})) - x_i x_j = 0.$$

Now multiply equation (9) by $\sigma_{ij}$ and sum to obtain

$$(10) \qquad E_{\mathrm{Sel}}\left(\sum_{i,j} \sigma_{ij} X_i X_j \Big| \hat{\boldsymbol{\Sigma}}(\mathbf{x})\right) - \sum_{i,j} \sigma_{ij} x_i x_j = 0.$$

Now, assume $\mathbf{x}^*$ satisfying the inequality (7) are the observed data. Then it follows from equation (7) that the newly defined random variable

$$(11) \qquad M^* \equiv \sum_{i,j} \sigma_{ij} X_i X_j - \sum_{i,j} \sigma_{ij} x_i^* x_j^* \geq 0,$$

with probability 1. Assume the likelihood given by equation (1) is bounded at the point $\mathbf{x}^*$. By continuity, the maximum likelihood estimate $\hat{\boldsymbol{\Sigma}}(\mathbf{x}^*)$ exists. Then it follows from equation (10) that

$$(12) \qquad E_{\text{Sel}}(M^*|\hat{\boldsymbol{\Sigma}}(\mathbf{x}^*)) = 0.$$

Therefore, $M^*$ is a non-negative random variable with mean zero, implying $M^* = 0$ with probability 1. However, $M^*$ is a continuous function of the continuous random vector $\mathbf{X}$ and is therefore equal to zero with probability zero. So, assuming the likelihood is bounded at $\mathbf{x}^*$ leads to a contradiction.

## APPENDIX B: PROOF OF COROLLARY 1

PROOF OF PART A.   Because the likelihood given by equation (3) is a smooth function of $\sigma$, standard calculus methods can be used to derive the MLE for $\sigma$. Again assuming that $\theta$ is known and differentiating the natural log of the likelihood given in (3), we get

$$(13) \qquad \frac{\partial}{\partial \sigma} \ln f_{\text{Sel}}(\mathbf{x}|\sigma, \theta) = -h + E_{\text{Sel}}(H|\sigma).$$

Define $g(\sigma) \equiv E_{\text{Sel}}(H|\sigma)$, then $g$ is a decreasing function. To show that $g$ is decreasing, we show that $g'(\sigma) < 0$ for all $\sigma$. It follows from equation (3) that

$$g(\sigma) = E_{\text{Sel}}(H|\sigma) = \frac{E_{\text{Neut}}(He^{-\sigma H})}{E_{\text{Neut}}(e^{-\sigma H})}$$

and

$$g'(\sigma) = \frac{-E_{\text{Neut}}(e^{-\sigma H})E_{\text{Neut}}(H^2 e^{-\sigma H}) + (E_{\text{Neut}}(He^{-\sigma H}))^2}{(E_{\text{Neut}}(e^{-\sigma H}))^2}$$

$$= -\text{Var}_{\text{Sel}}(H) < 0.$$

That is, as the selection intensity $\sigma$ grows, then homozygotes are increasingly disadvantaged. Thus, the mean homozygosity, $E_{\text{Sel}}(H|\sigma)$, will become smaller as the selection intensity, $\sigma$, grows. By equation (13), the maximum likelihood estimate for $\sigma$ will satisfy the equation

$$H = g(\hat{\sigma}).$$

To establish equation (4), we need to show that for every sequence of homozygosities converging to $1/k$ from above, the corresponding MLE for the selection intensity converges to infinity. Consider a monotone decreasing sequence of real numbers $\{a_n\}$ where $a_n \to 0$ as $n \to \infty$. Define $\sigma_n$ to be the solution to the equation

$$g(\sigma_n) = 1/k + a_n.$$

Since $g$ is a decreasing function, it follows by monotonicity that $\sigma_n$ must be an increasing sequence. Increasing sequences must either converge or diverge to infinity. Suppose for a moment that $\sigma_n \to \sigma^* < \infty$ as $n \to \infty$. Then by continuity of $g$, we have that $g(\sigma^*) = 1/k$. This implies that

(14) $$E_{\text{Sel}}(H - 1/k|\sigma^*) = 0.$$

However, we know that $H \geq 1/k$ with probability one. Therefore, it follows from equation (14) that $H - 1/k$ is a nonnegative random variable with mean zero. This implies that $H - 1/k$ is identically zero with probability one. This is a contradiction, since we know that, for any value of $\sigma$, $H$ is a continuous random variable whose distribution can be derived from equation (3) and so cannot be equal to $1/k$ with probability one. Therefore, $\sigma_n$ must go to infinity. $\square$

PROOF OF PART B. Note that $g(\sigma) = E(H|\sigma)$ is a decreasing function in $\sigma$ that is bounded below by $1/k$. Therefore, $\lim_{\sigma \to \infty} g(\sigma)$ must converge. Denote the limit by $b$. Note that $g^{-1}(b) = \infty$. By part (a), $g^{-1}(b)$ is the maximum likelihood estimate for $\sigma$ when $H = b$. Since the maximum likelihood estimate is finite for all $h > /1k$, then $b = 1/k$. Therefore, $E(H - 1/k|\sigma) = E(|H - 1/k| \, |\sigma)$ goes to zero as $\sigma \to \infty$. This implies that the $L_1$ norm of $H$ converges to $1/k$, which implies that $H$ converges to $1/k$ in probability. The conclusion of part (b) is established by noting that $H = 1/k$ if and only if $(X_1, X_2, \ldots, X_k) = (1/k, 1/k, \ldots, 1/k)$. $\square$

**Acknowledgments.** The authors would like to thank Craig Miller and Darin Rokyta for their helpful comments on the manuscript.

## REFERENCES

ALLISON, A. C. (1956). The sickle-cell and haemoglobin c genes in some African populations. *Am. Hum. Genet.* **21** 67–89.

CAVALLI-SFORZA, L. L. and BODMER, W. F. (1971). *The Genetics of Human Populations.* Dover, Mineola, NY.

DONNELLY, P., NORDBORG, M. and JOYCE, P. (2001). Likelihood and simulation methods for a class of nonneutral population genetics models. *Genetics* **159** 853–867.

EWENS, W. J. (2004). *Mathematical Population Genetics: I. Theoretical Introduction*, 2nd ed. Interdisciplinary Applied Mathematics **27**. Springer, New York. MR2026891

FEARNHEAD, P. and DONNELLY, P. (2001). Estimating recombination rates from population genetic data. *Genetics* **159** 1299–1318.

GENZ, A. and JOYCE, P. (2003). Computation of the normalization constant for exponentially weighted Dirichlet Distribution. *Comput. Sci. Statist.* **35** 557–563.

GILLESPIE, J. (1999). The role of population size in molecular evolution. *Theor. Popul. Biol.* **55** 145–156.

GRIFFITHS, R. C. and MARJORAM, P. (1997). An ancestral recombination graph. In *Progress in Population Genetics and Human Evolution, IMA Volumes in Mathematics and Its Applications.* (P. Donnelly and S. Tavaré, eds.) 257–270. Springer, Berlin. MR1493031

HARDING, R. M., FULLERTON, S. M. and GRIFFITHS, R. C. (1997). Archaic African and Asian lineages in the genetic ancestry of modern humans. *Am. Jour. Hum. Genet.* **60** 772–789.

HASTINGS, W. K. (1970). Monte Carlo sampling methods using Markov chains and their applications. *Biometrika* **57** 97–109.

JOYCE, P., GENZ, A. and BUZBAS, E. O. (2009). Efficient simulation methods for a class of nonneutral population genetics models. *Theor. Popul. Biol.* To appear.

JOYCE, P., KRONE, S. M. and KURTZ, T. G. (2003). When can one detect overdominant selection in the infinite-alleles model? *Ann. Appl. Probab.* **13** 181–212. MR1951997

METROPOLIS, N., ROSENBLUTH, A. W., ROSENBLUTH, M. N., TELLER, A. H. and TELLER, E. (1953). Equation of state calculations by fast computing machines. *J. Chem. Phys.* **21** 1087–1092.

NEUHAUSER, C. (1999). The Ancestral Graph and Gene Genealogy under Frequency-Dependent Selection. *Theor. Popul. Biol.* **56** 203–214.

NORDBORG, M. (2000). Linkage disequilibrium, gene trees and selfing: An ancestral recombination graph with partial self-fertilization. *Genetics* **154** 923–929.

NORMAN, P. J., COOK, M. A., CAREY, B. S., CARRINGTON, C. V. F., VERITY, D. H., HAMEED, K., RAMDATH, D. D., CHANDANAYINGYONG, D., LEPPERT, M., STEPHENS, H. A. F. and VAUGHAN, R. W. (2004). SNP haplotypes and allele frequencies show evidence for disruptive and balancing selection in the human leukocyte receptor complex. *Immuno-Genetics* **56** 225–237.

PADHUKASAHASRAM, B., MARJORAM, P., WALL, J. D., BUSTAMANTE, C. and NORDBORG, M. (2008). Exploring population genetic models with recombination using efficient forward-time simulations. *Genetics* **178** 2417–2427.

QIU, W. G., BOSLER, E., CAMPBELL, J., UGINE, E., WANG, I.,BENJAMIN, J. L. and DYKHUZEN, D. E. (1997). A population genetic study of Borrelia burgdorferi sensu stricto from eastern long island, New York, suggested frequency-dependent selection, gene flow and host adaptation. *Hereditas* **127** 203–216.

SNOW, R. W., GUERRA, C. A., NOOR, A. M., MYINT, H. Y. and HAY, S. I. (2005). The global distribution of clinical episodes of plasmodium falciparum malaria. *Nature* **434** 214–217.

WAKELEY, J. (2005). The limits of population genetics. *Genetics* **169** 1–7.

WATTERSON, G. A. (1977). Heterosis or neutrality? *Genetics* **85** 789–814. MR0504021

WRIGHT, S. (1949). Adaptation and selection. In *Genetics, Paleontology, and Evolution* (G. L. Jepson, G. G. Simpson and E. Mayr, eds.) 365–389. Princeton Univ. Press, Princeton.

DEPARTMENT OF MATHEMATICS,
BRINK HALL 300,
UNIVERSITY OF IDAHO,
MOSCOW, IDAHO 83844-1103
USA
E-MAIL: buzb2462@vandals.uidaho.edu
        joyce@uidaho.edu